\documentclass[nofootinbib,a4paper,aps,prd,10pt,superscriptaddress,showkeys,twocolumn]{revtex4}

\usepackage{graphicx}
\usepackage{amsfonts}
\usepackage{amssymb} 
\usepackage{amsmath}
\usepackage{hyperref}
\usepackage{natbib}
\usepackage{float}
\usepackage{dcolumn}

\usepackage{bm}
\usepackage{latexsym,color}

\newcommand{\bea}{\begin{eqnarray}}
\newcommand{\eea}{\end{eqnarray}}
\newcommand{\be}{\begin{equation}}
\newcommand{\ee}{\end{equation}}

\begin{document}

\title{Accretion disk luminosity around rotating naked singularities}

\author{Yergali~\surname{Kurmanov}}
\email[]{kurmanov.yergali@kaznu.kz}
\affiliation{National Nanotechnology Laboratory of Open Type, Almaty 050040, Kazakhstan.}
\affiliation{Al-Farabi Kazakh National University, Al-Farabi av. 71, Almaty 050040, Kazakhstan.}

\author{Kuantay~\surname{Boshkayev}}
\email[]{kuantay@mail.ru}
\affiliation{National Nanotechnology Laboratory of Open Type, Almaty 050040, Kazakhstan.}
\affiliation{Al-Farabi Kazakh National University, Al-Farabi av. 71, Almaty 050040, Kazakhstan.}
\affiliation{Institute of Nuclear Physics, Ibragimova, 1, Almaty, 050032, Kazakhstan}

\author{Talgar~\surname{Konysbayev}}
\email[]{talgar\_777@mail.ru}
\affiliation{National Nanotechnology Laboratory of Open Type, Almaty 050040, Kazakhstan.}
\affiliation{Al-Farabi Kazakh National University, Al-Farabi av. 71, Almaty 050040, Kazakhstan.}

\author{Marco Muccino}
\email[]{marco.muccino@lnf.infn.it}
\affiliation{Al-Farabi Kazakh National University, Al-Farabi av. 71, Almaty 050040, Kazakhstan.}
\affiliation{Universit\`a di Camerino, Divisione di Fisica, Via Madonna delle carceri 9, 62032 Camerino, Italy.}
\affiliation{ICRANet, Piazza della Repubblica 10, 65122 Pescara, Italy.}

\author{Orlando Luongo}
\email[]{orlando.luongo@unicam.it}
\affiliation{Al-Farabi Kazakh National University, Al-Farabi av. 71, Almaty 050040, Kazakhstan.}
\affiliation{Universit\`a di Camerino, Divisione di Fisica, Via Madonna delle carceri 9, 62032 Camerino, Italy.}
\affiliation{SUNY Polytechnic Institute, 13502 Utica, New York, USA.}
\affiliation{INAF, Osservatorio Astronomico di Brera, Milano, Italy.}
\affiliation{INFN, Sezione di Perugia, Perugia, 06123, Italy.}

\author{Ainur~\surname{Urazalina}}
\email[]{y.a.a.707@mail.ru}
\affiliation{National Nanotechnology Laboratory of Open Type,  Almaty 050040, Kazakhstan.}
\affiliation{Al-Farabi Kazakh National University, Al-Farabi av. 71, Almaty 050040, Kazakhstan.}
\affiliation{Institute of Nuclear Physics, Ibragimova, 1, Almaty, 050032, Kazakhstan}

\author{Anar~\surname{Dalelkhankyzy}}
\email[]{dalelkhankyzy.d@gmail.com}
\affiliation{Kazakh National Women’s Teacher Training University, Ayteke Bi, 99, Almaty 050040, Kazakhstan.}

\author{Farida~\surname{Belissarova}}
\email[]{farida.belisarova@kaznu.kz}
\affiliation{Al-Farabi Kazakh National University, Al-Farabi av. 71, Almaty 050040, Kazakhstan.}

\author{Madina~\surname{Alimkulova}}
\email[]{m.alimkulova@mail.ru}
\affiliation{Al-Farabi Kazakh National University, Al-Farabi av. 71, Almaty 050040, Kazakhstan.}

\date{\today}

\begin{abstract}
We explore circular geodesics of neutral test particles in the gravitational field of a rotating deformed mass. The geometry around this source is described by the Quevedo-Mashhoon solution, which corresponds to a naked singularity. To this end, we compute the orbital parameters of test particles in the equatorial plane, such as the angular velocity $\Omega$, energy $E$ and angular momentum $L$, per unit mass. We also numerically estimate the radius of the innermost stable circular orbit $r_{ISCO}$. In addition, we consider a simple model for the disk's radiative flux, differential luminosity, and spectral luminosity, employing the well-known Novikov-Page-Thorne model. We mainly focus on the flux of the accretion disk around several rotating naked singularities possessing the same mass and quadrupole moment but different rotation and deformation parameters. We analyze the possibility of distinguishing Kerr black holes from rotating naked singularities. The astrophysical implications of the results obtained are discussed.
\end{abstract}


\maketitle
 
\section{Introduction}

Gravitational collapse and associated spacetime singularities are fundamental to several key aspects of relativistic astrophysics and are important in gravitation theory \cite{1939PhRv...56..455O,1999PThPS.136...45K,2007arXiv0710.1818H,2014shst.book..409J}. When massive stars reach the final stages of their nuclear burning cycles, they undergo continuous collapse due to their own gravity, forming ultra-compact structures like black holes (BHs) and spacetime singularities.  Numerous observational studies provide strong evidence supporting the existence of astrophysical BH candidates, such as the detection of gravitational waves from binary BH mergers \cite{Abbott, Abbott2018, Abbott2016, Abbott2017, AbbottPhysRevLett, Abbott2017ApJ} and the imaging of shadows formed by supermassive objects located at the centers of the M87 and Milky Way galaxies \cite{EventHorizonTelescope:2019dse, 2019ApJS..243...26P, 2019ApJ...875L...2E,2019ApJ...875L...3E,2019ApJ...875L...5E, EventHorizonTelescope:2019ggy, EventHorizonTelescope:2022wkp, 2022ApJ...930L..13E,2022ApJ...930L..15E,2022ApJ...930L..16E,2022ApJ...930L..17E,2022ApJ...930L..20G, 2022ApJ...930L..21B, 2022ApJ...930L..18F}.

In addition to BHs, different class of compact objects so-called naked singularities (NSs) represent another potential outcome of gravitational collapse \cite{1993PhRvL..70....9C,1995PhRvD..51.4168B,2018AnHP...19..619A}. NSs have become a focal point of considerable theoretical research in the realm of GR. In the literature, several works have explored the emergence of accreting systems around NSs \cite{2011PhRvD..83b4021P,2014PhRvD..90b4035V,2015EPJC...75..451S,2020EPJC...80.1017G,2023MNRAS.523.4615V,2024PhRvL.133x1401K,2024CQGra..41f5004T,2025arXiv250103178C} and various properties of accretion disks around BHs \cite{2021PhRvD.104h4009B,2025PDU....4701743U,2022PhRvD.106d4036T,2021CQGra..38c5012G,2024arXiv241208447L,2024arXiv241110315Z,2024NewA..11102249U,2024EPJC...84.1075W,2024Ap&SS.369...96H,2024JCAP...08..041Z,2025arXiv250101018L}.

One of the perplexing features of BHs is the presence of singularities at their cores predicted by exact solutions to Einstein's equations. However, the inability of GR to physically describe singularities highlights its limitations. A potential solution to this problem involves the construction of so-called regular BH solutions \cite{1998PhRvL..80.5056A,1999PhLB..464...25A}. These models describe spacetimes with event horizons, but avoid the occurrence of singularities \cite{2023PhRvD.108d4063B,2024PDU....4501525S,2024CQGra..41l5005G}. The concept of a regular BH was first proposed by Bardeen \cite{Bardeen1968}.

Accreting BHs range in size from stellar-mass BHs in X-ray binaries \cite{Verbunt, Remillard} to supermassive BHs in galactic centers \cite{Kormendy}. Supermassive BHs with masses from millions to tens of billions of solar masses reside in the centers of almost all galaxies \cite{Gebhardt,Walsh,GC}. In various astrophysical scenarios including galactic nuclei, binary star interactions, and young stellar objects, accretion disks are vital components because they allow for the detection of radiation from matter orbiting a compact object. Recent advances in astronomical techniques have dramatically improved the accuracy of observing these disks \cite{2019ApJ...870..123E,2022MNRAS.511.5346S,2024ApJS..272...26S}, leading to more precise measurement of essential properties such as temperature and luminosity \cite{2023PhRvD.107d3032D,2024arXiv240807762A,2025PhRvD.111b4013B,2024IJGMM..2140019C}.

Analyzing the emission spectra from these disks can offer insight into the nature of the accreting central object. As a BH accretes more matter, it produces sufficient momentum and emits energy into its environment, expelling gas from the galaxy and effectively cutting off its fuel supply \cite{Murray,Silk,King}. Substantial evidence supports the presence of rotating BHs and event horizons, provided by the study of the luminosity spectra of accreting BHs.

Certain characteristics of accretion disks are influenced by geometry \cite{2016gac..conf..185A} and, accordingly, accretion disks can be used to fix constraints on geometry through observational data \cite{abramowicz}. As a direct example, it has been studied the motion of test particles in the gravitational field of a Schwarzschild black hole surrounded by dark matter with a spherically symmetric distribution, characterized by nonzero isotropic, anisotropic, and tangential pressures \cite{2020MNRAS.496.1115B, 2022ApJ...925..210K, 2022ApJ...936...96B}. Additionally, it has been investigated the thermodynamic properties of thin accretion disks in exact regular spacetimes derived from nonlinear electrodynamics (NLED), which may be relevant for modeling singularity-free compact objects \cite{2024PDU....4601566K}. Moreover, even the thermodynamic and spectral properties of accretion disks around rotating Hayward black holes \cite{2024EPJC...84..230B} and rotating Bardeen black holes \cite{kurmanov2024radiative} have been worked out, as well as the motion of test particles in circular orbits within the Hartle-Thorne (HT) geometry \cite{2024EPJP..139..273B}, conducting a detailed analysis of the accretion disk’s flux, differential luminosity, and spectral luminosity \cite{2021PhRvD.104h4009B}. 

In this work, we analyze the radiative flux emitted from the surroundings of compact objects in the traditional theory of accretion outlined in \cite{novikov1973,page1974}, which can be adjusted to fit specific spacetime symmetries. We focus on exploring a limiting class of solutions derived by Quevedo and Mashhoon \cite{Quevedo1991} that, in the most general situation, exhibit infinite sets of gravitational and electromagnetic multipole moments. The Quevedo-Mashhoon (QM) solution is an exact external metric that describes the gravitational field of a rotating deformed mass \cite{Quevedo1985,Quevedo1986,Quevedo1989,Quevedo1990}, which belongs to the Weyl–Lewis–Papapetrou class \cite{Weyl1917, Lewis1932, Papapetrou1953}. We confine ourselves to a particular case of a metric that includes the mass parameter $m$, the quadrupole parameter $q$, and the rotation parameter $a$, which is a generalization of the Kerr metric \cite{Kerr1963}. For vanishing $a$, the particular case of the QM metric reduces to the well-corroborated Erez-Rosen (ER) solution which describes deformed objects (naked singularities)  \cite{ER,Bini2009}. It is interesting to note that there are some solutions, whose physical characteristics are close to the QM metric. For example, in ~\cite{1992CQGra...9.2477M} the authors explored two asymptotically flat metrics that describe the superposition of the Kerr solution with a static vacuum Weyl field, distinguished by their angular momentum distributions. An asymptotically flat solution of the Einstein vacuum field equations which simplifies the Schwarzschild spacetime in the static case and accurately describes the external gravitational field of a spinning mass due to an infinite set of arbitrary multipole moments was examined in \cite{1990CQGra...7..779C}. Hence, the circular geodesics of test particles in the QM space-time are here examined using standard techniques of GR. The radius of the innermost stable circular orbit $r_{ISCO}$ and the flux of electromagnetic radiation in the accretion disk are calculated and analyzed. Then, the differential and spectral luminosities of the accretion disk are estimated. The purpose of this paper is to study the flux of electromagnetic radiation and the accretion disk's luminosity in the QM spacetime, when the mass and quadrupole moment of the source are fixed. Thus, we assess the feasibility of differentiating rotating BHs from rotating NSs, which occasionally are dubbed BH mimickers.
 
The paper is organized as follows. In Sect. \ref{sez2}, we present the QM spacetime and review its main physical properties. In Sect. \ref{sez3}, we examine circular orbits in the QM spacetime and review the definitions of flux, differential luminosity, and spectral luminosity within the framework of the Novikov–Page–Thorne model, while Sect. \ref{sez4} is devoted to numerical results of the spectral properties of the accretion disks. Finally, in Sect. \ref{sez5}, we discuss our results and comment on possible perspectives. Throughout the paper, we make use of geometrized units by setting $G=c=1$.

\section{Particle motion in the QM metric}\label{sez2}

In this study, we work with the particular case of the QM metric, described by the following line element \cite{Quevedo1985,Quevedo1986,Quevedo1989,Quevedo1990,2018RSOS....580640F}
\begin{eqnarray}
ds^2&=-f\left(dt-\omega d\phi\right)^{2}+\frac{1}{f}\Bigg[e^{2\gamma}\left(d\theta^{2}+\frac{dr^2}{r^{2}-2mr+a^{2}}\right)\nonumber\\
&\times \Big[\left(m-r\right)^{2}-\left(m^2-a^{2}\right)\cos^{2}\theta\Big]\nonumber\\
&+\left(r^{2}-2mr+a^{2}\right)\sin^{2}\theta d\phi^{2}\Bigg],
\label{eq:metric}
\end{eqnarray}
where $f$, $\omega$ and $\gamma$ depend solely on $r$ and $\theta$. These functions take the form $[x=\left(r-m\right)/\sigma,  \ \  y=\cos\theta]$
\begin{eqnarray}\label{eq:2}
f&=&\frac{\Tilde{R}}{\Tilde{L}}e^{-2q\delta P_{2}Q_{2}},\nonumber\\
\omega &=&-2a-2\sigma\frac{\Tilde{M}}{\Tilde{R}}e^{2q\delta P_{2}Q_{2}},\nonumber\\
e^{2\gamma}&=&\frac{1}{4}\left(1+\frac{m}{\sigma}\right)^{2}\frac{\Tilde{R}}{(x^{2}-1)^{\delta}}e^{2\delta^{2}\hat{\gamma}},
\end{eqnarray}
denoting
\begin{eqnarray}\label{eq:3b}
\Tilde{R}&=&a_{+}a_{-}+b_{+}b_{-},\nonumber\\
\Tilde L&=&a^{2}_{+}+b^{2}_{+},
\end{eqnarray}
\begin{eqnarray}
\Tilde{M}&=(x+1)^{\delta-1}\Big[x(1-y^{2})(\lambda+\eta)a_{+}\nonumber\\
&+y(x^{2}-1)(1-\lambda\eta)b_{+}\Big],
\label{eq:3}
\end{eqnarray}
\begin{eqnarray}
\hat{y}&=&\frac{1}{2}(1+q)^{2}\ln\frac{x^{2}-1}{x^{2}-y^{2}}+2q(1-P_{2})Q_{1}\nonumber\\
&+&q^{2}(1-P_{2})\Big[(1+P_{2})(Q^{2}_{1}-Q^{2}_{2})\nonumber\\
&+&\frac{1}{2}(x^{2}-1)(2Q^{2}_{2}-3xQ_{1}Q_{2}+3Q_{0}Q_{2}-Q^{'}_{2})\Big].
\label{eq:4}
\end{eqnarray}
In this context, $P_{l}(y)$ and $Q_{l}(x)$ correspond to the Legendre polynomials of first and second kind, respectively. Moreover,
\begin{eqnarray}
a_{\pm}&=&(x \pm 1)^{\delta -1}\Big[x(1-\lambda \eta)\pm(1+\lambda\eta)\Big],
\label{eq:5}
\end{eqnarray}
\begin{eqnarray}
b_{\pm}&=&(x \pm 1)^{\delta -1}\Big[y(\lambda + \eta)\mp (\lambda-\eta)\Big],
\label{eq:6}
\end{eqnarray}
with
\begin{eqnarray}
\lambda &=&\alpha \left(x^{2}-1\right)^{1-\delta}\left(x+y\right)^{2\delta -2}e^{2q\delta \delta_{+}},
\label{eq:7}
\end{eqnarray}
\begin{eqnarray}
\eta &=&\alpha \left(x^{2}-1\right)^{1-\delta}\left(x-y\right)^{2\delta -2}e^{2q\delta \delta_{-}},
\label{eq:88}
\end{eqnarray}
\begin{eqnarray}
\delta_{\pm}&=&\frac{1}{2}\ln \frac{\left(x\pm y \right)^{2}}{x^{2}-1}+\frac{3}{2}\left(1-y^{2}\mp xy\right)\nonumber\\
&+& \frac{3}{4}\Big[x\left(1-y^{2}\right)\mp y\left(x^{2}-1\right)\Big]\ln \frac{x-1}{x+1},
\label{eq:8}
\end{eqnarray}
$\alpha$ and $\sigma$ are constants defined as
\begin{align}
\alpha &=\frac{\sigma-m}{a}, &\qquad \sigma &=\sqrt{m^{2}-a^{2}}. 
\end{align}
By computing the Geroch-Hansen moments \cite{Geroch,Hansen}, the physical significance of the parameters entering this metric may be examined in an invariant way:
\begin{align}
M_{2k+1} &=J_{2k}=0, &\quad k &=0,1,2,..., 
\end{align}
\begin{align}
M_{0} &=m+\sigma\left(\delta-1\right), &\quad J_{1} &=ma+2a\sigma\left(\delta-1\right), 
\end{align}
\begin{eqnarray}
M_{2}&=&-ma^{2}+\frac{2}{15}q\sigma^{3}-\frac{1}{15}\sigma\left(\delta-1\right)[45m^{2}\nonumber\\
&+&15m\sigma\left(\delta-1\right)-\left(30+2q+10\delta-5\delta^{2}\right)\sigma^{2}],
\label{M2}
\end{eqnarray}
\begin{eqnarray}
J_{3}&=&-ma^{3}+\frac{4}{15}aq\sigma^{3}-\frac{1}{15}a\sigma\left(\delta-1\right)[60m^{2}\nonumber\\
&+&45m\sigma\left(\delta-1\right)-2\sigma^{2}\left(15+2q+10\delta-5\delta^{2}\right)],
\label{J3}
\end{eqnarray}

The reflection symmetry with respect to the equatorial plane implies that the odd gravitoelectric $(M_{n})$ and even gravitomagnetic $(J_{n})$ multipole moments  vanish. In the case where $\delta=1$, $m = M$ represents the body's total mass, $q$ quantifies the deviation from spherical symmetry, and $a$ reports the specific angular momentum. Consequently, all higher multipole moments are uniquely determined by the variables $m$, $a$, and $q$. Generally speaking, it is evident that the source's quadrupole moment is related to the Zipoy-Voorhees parameter.

\section{Circular Orbits and Accretion Disks in the QM Spacetime}\label{sez3}

In order to examine the observational characteristics of thin accretion disks, we begin our analysis with the study of motion of neutral massive test particles composing the disk. These particles are considered to follow circular paths within the equatorial plane, where  $\theta = \pi/2$. 

For circular orbits in the equatorial plane ($\theta=\pi/2$), with conditions $\dot{r}=\dot{\theta}=\ddot{r}=0$ the geodesic equation leads to the following relation for the angular velocity of the test particles \cite{2017Bambi}
\begin{equation}
\Omega_{\pm}=\frac{-g_{t\phi,r}\pm\sqrt{(g_{t\phi,r})^2-g_{tt,r}g_{\phi\phi,r}}}{g_{\phi\phi,r}},
\label{11}
\end{equation}
The energy ${E}$ per unit mass and the specific angular momentum ${L}$ per unit mass of the test particles moving in circular orbits as given by
\begin{subequations}
\begin{align}
{{E}}&=-\frac{g_{tt}+g_{t\phi}\Omega}{\sqrt{-g_{tt}-2g_{t\phi}\Omega-g_{\phi\phi}\Omega^2}},
\label{12}\\
{{L}}&=\frac{g_{t\phi}+g_{\phi\phi}\Omega}{\sqrt{-g_{tt}-2g_{t\phi}\Omega-g_{\phi\phi}\Omega^2}}.
\label{13}
\end{align}
\end{subequations}
\begin{figure*}[t]
\begin{minipage}{0.49\linewidth}
\center{\includegraphics[width=0.97\linewidth]{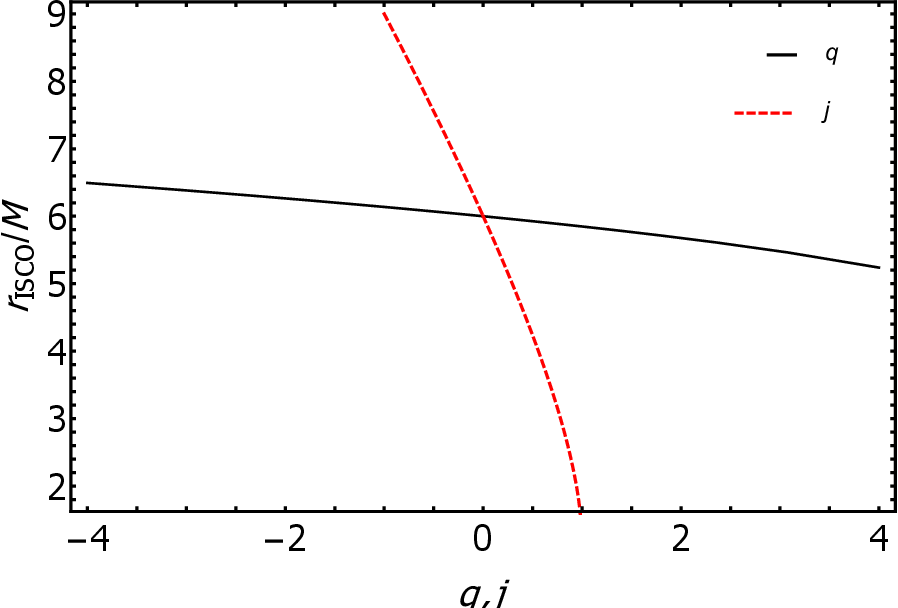}}\\ 
\end{minipage}
\hfill 
\begin{minipage}{0.50\linewidth}
\center{\includegraphics[width=0.97\linewidth]{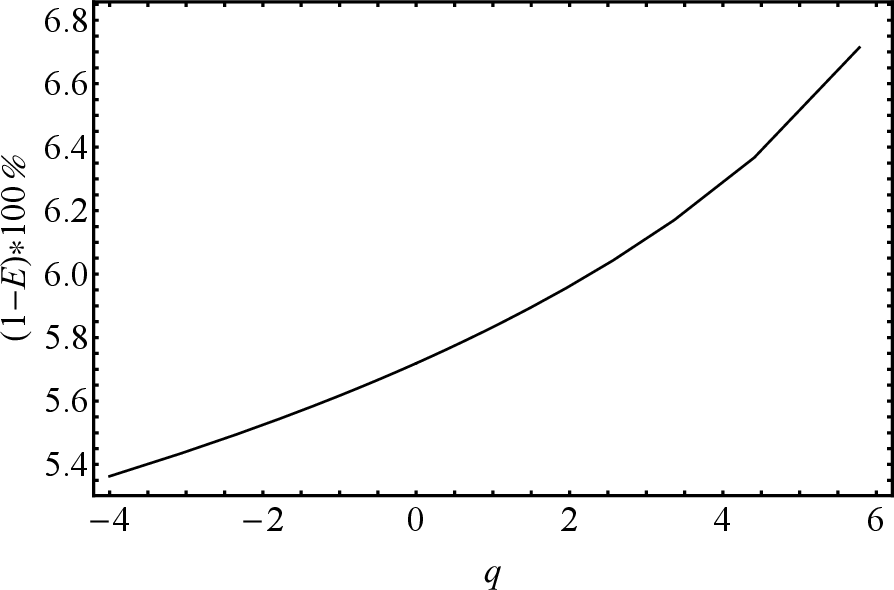}}\\ 
\end{minipage}
\caption{Left panel: The ER-space-time's  ISCO radii as compared to the Kerr space-time's ISCO. Right panel: Radiative efficiency for the ER metric.}
\label{fig:riscoqj}
\end{figure*}

\begin{figure*}[ht]
\begin{minipage}{0.49\linewidth}
\center{\includegraphics[width=0.97\linewidth]{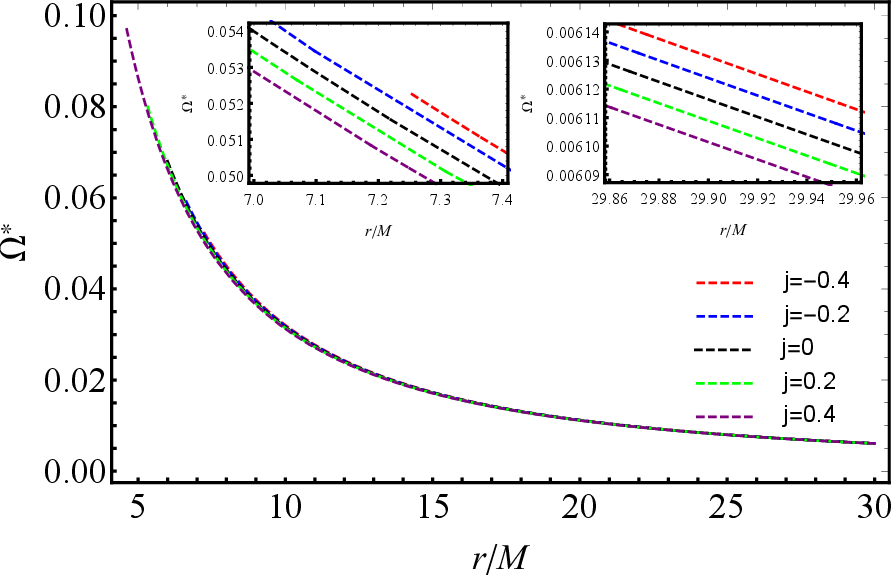}\\ }
\end{minipage}
\hfill 
\begin{minipage}{0.50\linewidth}
\center{\includegraphics[width=0.97\linewidth]{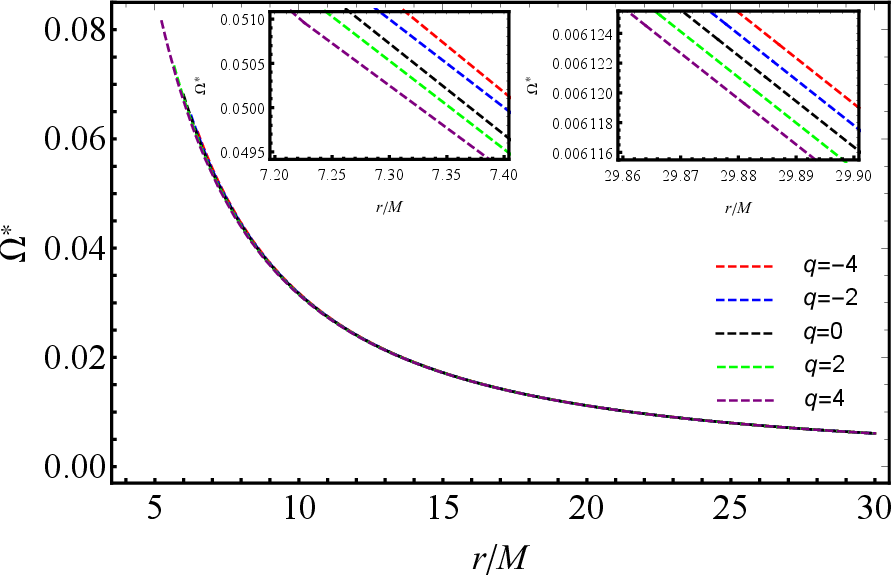}\\ }
\end{minipage}
\caption{Left panel: Angular velocity of test particles versus radial distance $r$ normalized in units of total mass $M$ in the QM metric with $q=0$. Right panel: Angular velocity of test particles versus radial distance $r$ normalized in units of total mass $M$ in the QM metric with $j=0$.}
\label{fig:omega}
\end{figure*}

The radius of the innermost stable circular orbit $r_{ISCO}$ is defined as $dL/dr = 0$ \cite{2016PhRvD..93b4024B}. So, for the ER-metric $r_{ISCO}$ calculated numerically and the radius of the ISCO for the Kerr metric derived as \cite{2021PhRvD.104h4009B}.ISCO radii for the ER-metric in units of total mass M as a function of the quadrupole parameter $q$ compared to the ISCO for the Kerr spacetime as a function of the dimensionless angular momentum $j$, shown on the left panel of Fig. \ref{fig:riscoqj}.

By examining the energy dissipation of a test particle moving from infinity to the disk's inner edge, we calculate the efficiency of converting mass to radiation, $\epsilon$. All photons released from the disk's surface are thought to be capable of reaching infinity. As the radius approaches infinity, the particle's energy tends to unity, or $E_\infty\approx1$. Thus,

\begin{eqnarray}
\label{eq.028}
\epsilon=\frac{E_\infty-E_{isco}}{E_\infty}\approx1-E_{isco},
\end{eqnarray}

In the right panel of Fig. \ref{fig:riscoqj}, we compute the accretion disk efficiency in the ER metric. In particular, we evaluate the efficiency in converting mass into radiation and observe that for $q < 0$, the efficiency in the ER metric is lower than that obtained within the Schwarzschild case. Conversely, for $q > 0$, it looks higher. Further, an accretion disk around a Schwarzschild BH converts matter into radiation with an efficiency of 5.72\%, following the Schwarzschild efficiency formula, namely $1 - E_{ISCO} = 0.0572 $.

In order to study the luminosity and spectrum of the accretion disk in the QM metric, we follow the model proposed by Novikov-Thorne and Page-Thorne in \cite{novikov1973, page1974}.

We use Refs. \cite{2021PhRvD.104h4009B,2024EPJC...84..230B,2024EPJP..139..273B,2024PDU....4601566K} to calculate the radiative flux $\mathcal{F}$ (i.e. the energy radiated per unit area per unit time) emitted by the accretion disk, the differential luminosity (i.e. the energy per unit time that reaches an observer at infinity) $\mathcal{L}_{\infty}$ which can be estimated from the flux $\mathcal{F}$, the spectral luminosity distribution observed at infinity $\mathcal{L}_{\nu,\infty}$. Thus, the radiative flux is given by
\begin{equation}\label{eq:flux}
\mathcal{F}(r)=-\frac{\dot{{\rm m}}}{4\pi \sqrt{-g}} \frac{\Omega_{,r}}{\left(E-\Omega L\right)^2 }\int^r_{r_{ISCO}} \left(E-\Omega L\right) L_{,\tilde{r}}d\tilde{r},
\end{equation}
where $\dot{{\rm m}}$ is the mass accretion rate of the disk, which is assumed to be unity for simplicity, and $g$ is the metric tensor determinant of the three-dimensional sub-space ($t,r,\varphi$), (i.e. $\sqrt{-g}=\sqrt{-g_{rr}(g_{tt}g_{\varphi\varphi}-g_{t\varphi}^2}$) \cite{2012ApJ...761..174B}.
In addition to $\mathcal{F}$, the differential luminosity (i.e. the energy per unit time that reaches an observer at infinity) $\mathcal{L}_{\infty}$ is another physical quantity of astrophysical interest\cite{novikov1973, page1974}
\begin{equation}\label{eq:difflum}
\frac{d\mathcal{L}_{\infty}}{d\ln{r}}=4\pi r \sqrt{-g}E \mathcal{F}(r).
\end{equation}
Moreover, in practice, we measure the spectrum of the light emitted as a function of the frequency. Therefore, it is worth considering the spectral luminosity distribution observed at infinity $\mathcal{L}_{\nu,\infty}$. Under this assumption, the black body emission from the accretion disk $\mathcal{L}_{\nu,\infty}$ is given by \cite{2020MNRAS.496.1115B}
\begin{equation} \label{eq:speclum}
\nu \mathcal{L}_{\nu,\infty}=\frac{60}{\pi^3}\int^{\infty}_{r_{ISCO}}\frac{\sqrt{-g }E}{m^2}\frac{(u^t y)^4}{\exp\left[u^t y/\mathcal{F}^{*1/4}\right]-1}dr,
\end{equation}
where $u^t$ is the contra variant time component of the four-velocity, defined by
\begin{equation}\label{eq:sample7}
u^t(r)=\frac{1}{\sqrt{-g_{tt}-2\Omega g_{t\varphi}-\Omega^2 g_{\varphi \varphi}}},
\end{equation}
and $y=h\nu/kT_*$, $h$ is the Planck constant, $\nu$ is the frequency of the emitted radiation, $k$ is the Boltzmann constant, $T_*$ is the characteristic temperature as defined from the Stefan-Boltzmann law as $\sigma T_*^{4}=\dot{{\rm m}}/4\pi m^2$ with $\sigma$ being the Stefan-Boltzmann constant.
Notice that in order to keep the argument of the exponential dimensionless we have normalized the flux with respect to the mass $m$ and defined $\mathcal{F}^*(r)=m^2\mathcal{F}(r)$.

\begin{figure*}[ht]
\begin{minipage}{0.49\linewidth}
\center{\includegraphics[width=0.97\linewidth]{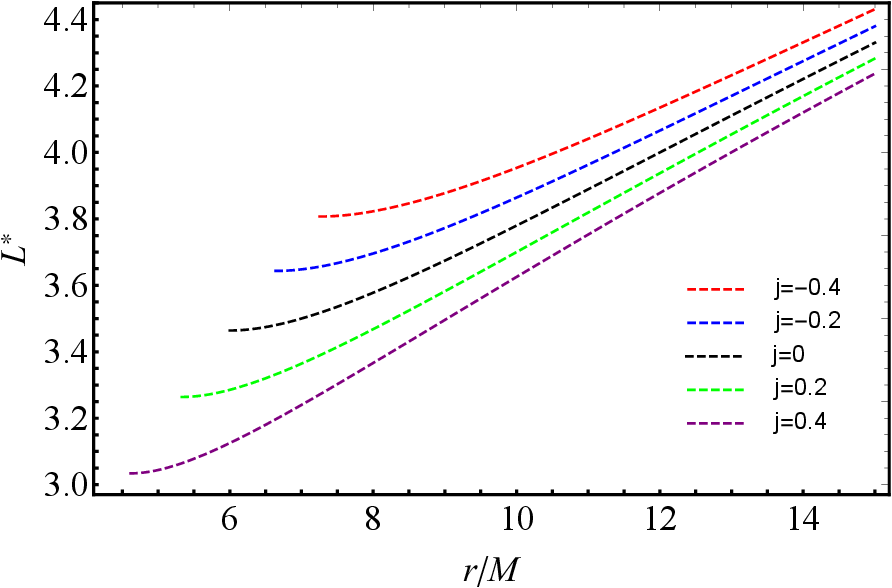}\\ } 
\end{minipage}
\hfill 
\begin{minipage}{0.50\linewidth}
\center{\includegraphics[width=0.97\linewidth]{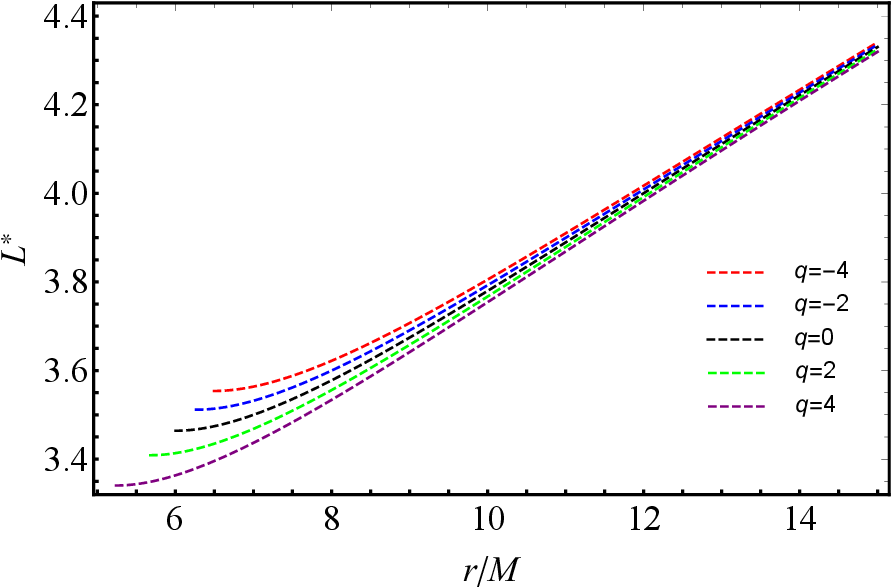}\\ }
\end{minipage}
\caption{Angular momentum $L^*$ of test particles versus radial distance $r$ normalized in units of total mass $M$ in the QM spacetime for $q=0$ with different values of $j$ (left panel) and for $j=0$ with different values of $q$ (right panel).}
\label{fig:angmom}
\end{figure*}
\begin{figure*}[ht]
\begin{minipage}{0.49\linewidth}
\center{\includegraphics[width=0.97\linewidth]{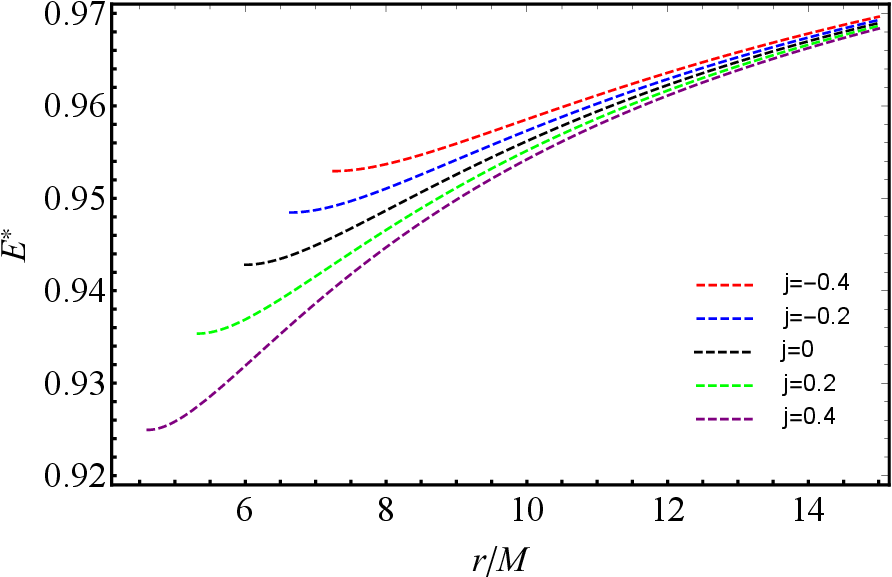}\\ } 
\end{minipage}
\hfill 
\begin{minipage}{0.50\linewidth}
\center{\includegraphics[width=0.97\linewidth]{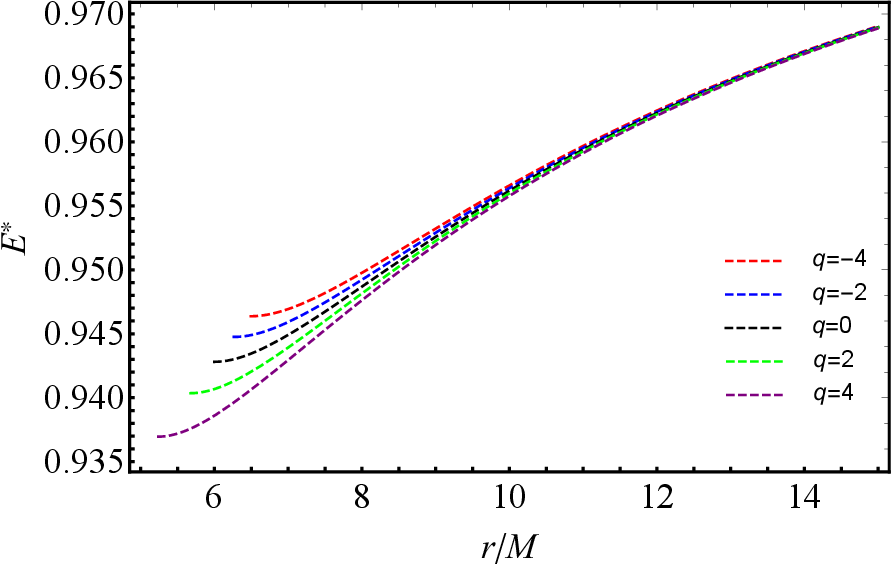}\\ }
\end{minipage}
\caption{Energy $E^*$ of test particles versus radial distance $r$ normalized in units of total mass $M$ for $q=0$ with different values of $j$ (left panel) and for $j=0$ with different values of $q$ (right panel).}
\label{fig:energy}
\end{figure*}
\begin{figure*}[ht]
\begin{minipage}{0.49\linewidth}
\center{\includegraphics[width=0.97\linewidth]{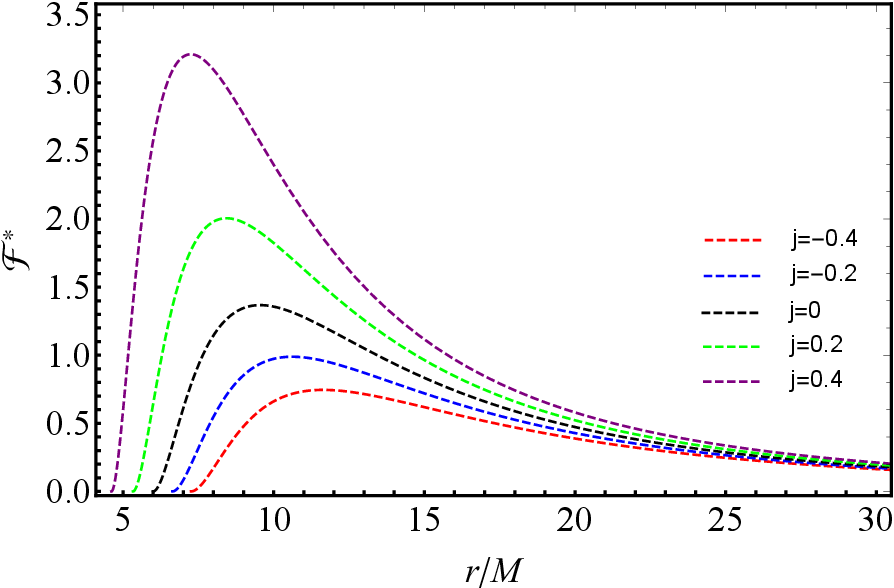}\\ } 
\end{minipage}
\hfill 
\begin{minipage}{0.50\linewidth}
\center{\includegraphics[width=0.97\linewidth]{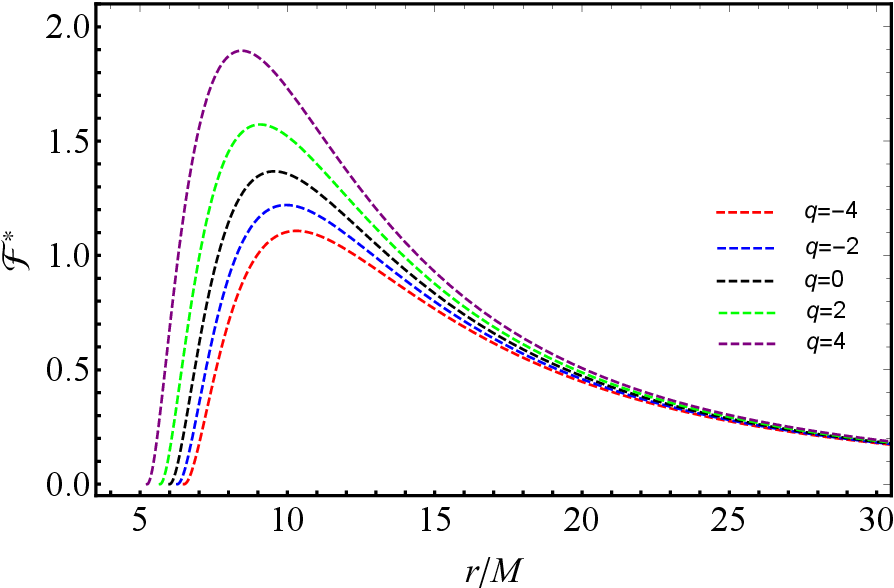}\\ }
\end{minipage}
\caption{Radiative flux $\mathcal{F}^*$ multiplied by $10^5$ of the accretion disk versus radial distance $r$ normalized in units of total mass $M$.  $\mathcal{F}^*$ in the  QM spacetime for $q=0$ with different values of $j$ (left panel) and for $j=0$ with different values of $q$ (right panel).}
\label{fig:flux}
\end{figure*}
\begin{figure*}[ht]
\begin{minipage}{0.49\linewidth}
\center{\includegraphics[width=0.97\linewidth]{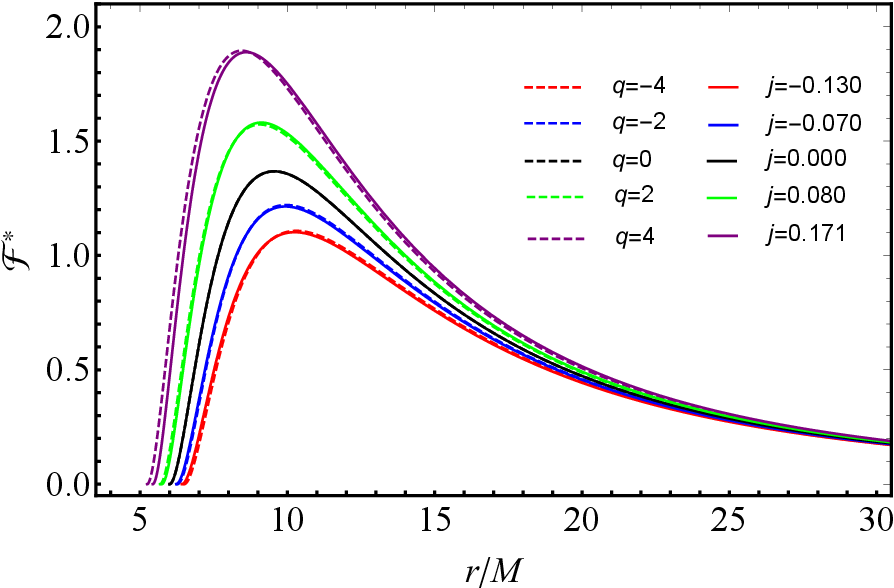}\\ } 
\end{minipage}
\hfill 
\begin{minipage}{0.50\linewidth}
\center{\includegraphics[width=0.97\linewidth]{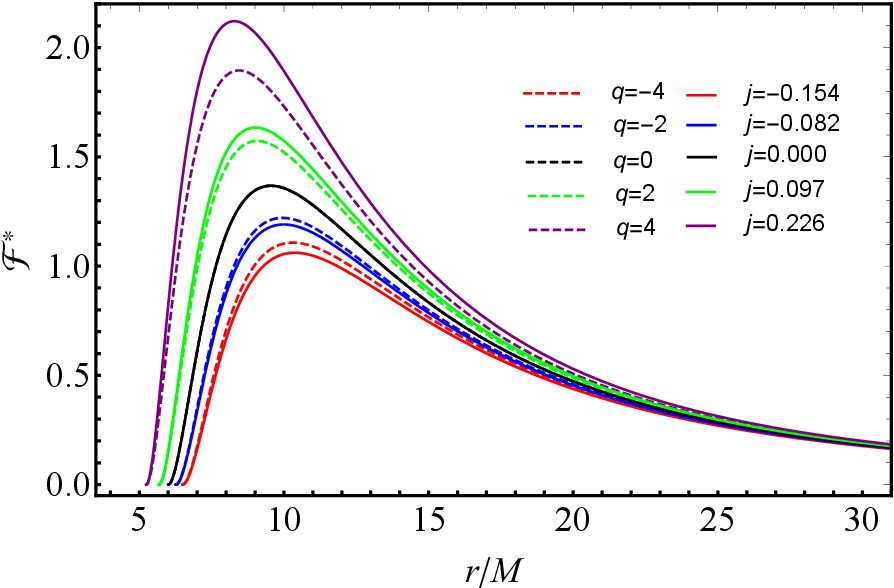}\\ }
\end{minipage}
\caption{Radiative flux $\mathcal{F}^*$ multiplied by $10^5$ of the accretion disk versus normalized radial distance $r/M$ in the ER and Kerr spacetimes. \textcolor{blue}{Left panel: The flux in ER metric for different values of $q$ mimicking the flux in the Kerr metric for values of $j$. Right panel: The flux $\mathcal{F}^*$ in the Kerr and ER metrics for fixed $r_{ISCO}$.}}
\label{fig:fluxisco}
\end{figure*}
Eq.~\eqref{M2} for $\delta=1$ and $m=M=1$ with the spin parameter $j=a/m$ becomes $M_2=-j^2+(2/15)q(1-j^2)^{3/2}$. Using this formula for fixed $M_{2}=\left[-0.1,-0.3,-0.5\right]$ and varying $j$  we find different values of $q_1$, $q_{2}$ and $q_{3}$ (see Tab.~\ref{tab:qjM} for details). 

\begin{table}[ht]
\begin{center}
\caption{The different values of the quadrupole parameters $q_{0}$, $q_{1}$, $q_{2}$ and $q_{3}$ for various values of fixed $M_{2}$, $j$, and $M=1$.}
\vspace{3 mm}
\label{tab:qjM}
\begin{tabular}{ccccc}
\hline
\hline
$ j $  &  $q_{0}$   & $q_{1}$        & $q_{2}$        & $q_{3}$  \\
       & ($M_2 = 0$) & ($M_2 = -0.1$) & ($M_2 = -0.3$) & ($M_2 = -0.5$) \\
\hline
-0.154 &     0.178         &  -0.572        &     -2.072     &  -3.572  \\   
-0.082 &     0.050         &  -0.699        &     -2.199     &  -3.699  \\
0      &     0             &  -0.750        &     -2.250     &  -3.750  \\
0.097  &     0.071         &  -0.679        &     -2.179     &  -3.679  \\
0.226  &     0.383         &  -0.367        &     -1.867     &  -3.367  \\ 
\hline
\hline
  \end{tabular}
  \end{center}
\end{table}
\begin{figure*}[ht]
\begin{minipage}{0.49\linewidth}
\center{\includegraphics[width=0.97\linewidth]{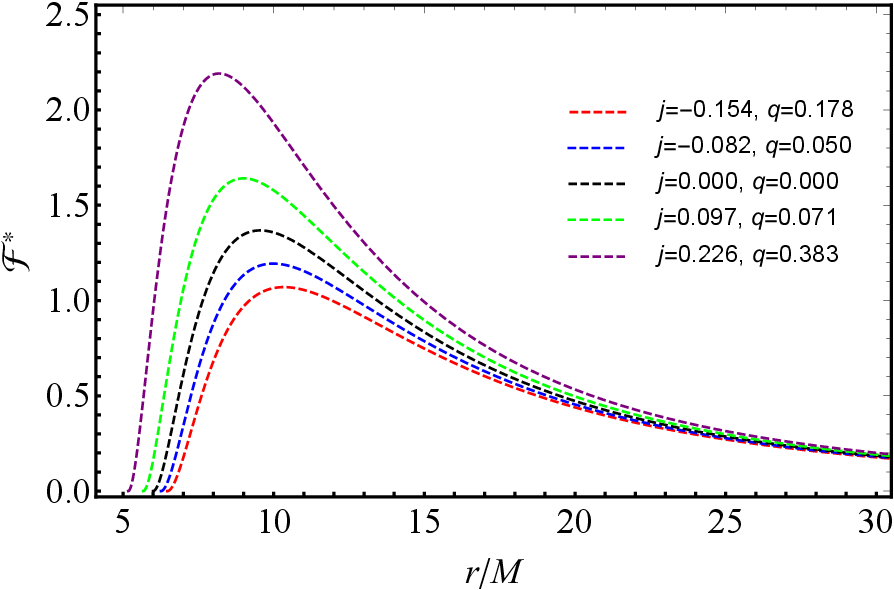}\\ } 
\end{minipage}
\hfill 
\begin{minipage}{0.50\linewidth}
\center{\includegraphics[width=0.97\linewidth]{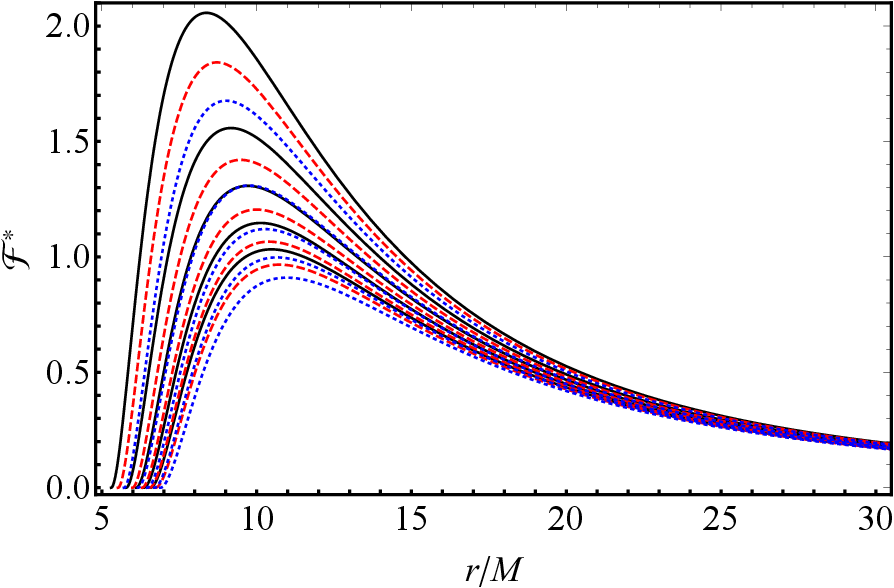}\\ }
\end{minipage}
\caption{Radiative flux $\mathcal{F}^*$ multiplied by $10^5$ of the accretion disk versus normalized radial distance $r/M$. Left panel: $\mathcal{F}^*$ in the QM spacetime with different values of $j$ and $q$. Right panel: $\mathcal{F}^*$ in the QM spacetime with different values of $j$ and the values of $q_{1}$ corresponds to $M_{2}=-0.1$ (black  curves), the values of $q_{2}$ corresponds to $M_{2}=-0.3$ (red dashed curves) and the values of $q_{3}$ corresponds to $M_{2}=-0.5$ (blue dotted curves) ( see in Tab. \ref{tab:qjM}).}
\label{fig:fluxM}
\end{figure*}

\section{Numerical results}\label{sez4}

We here present our main findings, regarding the orbital parameters of neutral massive test particles, including angular velocity, angular momentum, and energy in circular orbits. These quantities will be used to numerically evaluate the flux, as well as the differential and spectral luminosities of accretion disks in QM spacetime.  

Precisely, Fig.~\ref{fig:omega} shows the orbital angular velocity, $\Omega^*(r) = M \Omega$, of test particles as a function of the normalized radial coordinate, $r/M$, in the QM metric. 

In the left panel, we set $q = 0$ albeit  allowing $j$ to vary arbitrarily, corresponding to the Kerr metric. In the right panel, we fix $ j = 0$, varying $q$, corresponding to the ER spacetime. As observed in the left panel, for different negative (positive) values of $j$, the angular velocity is always higher (lower) than in the Schwarzschild case, say $j = 0$, represented by the black dashed curve. Similarly, in the right panel, we observe the same trend, but for negative and positive values of $q$ with $j = 0$ fixed.

In Fig.~\ref{fig:angmom} we constructed the dimensionless orbital angular momentum $L^*(r)$ of the test particles as a function of the normalized radial coordinate $r/M$ in the QM metric. The left panel corresponds to fixed $q = 0$ and arbitrary $j$, while the right panel represents fixed $j = 0$ and arbitrary $q$.  In the left panel, the curves of $L^*(r)$ for $j>0$ ($j<0$) are shown to be lower (higher) than in the Schwarzschild case (black dashed curve). In the right panel, it can be seen that the curves of $L^*(r)$ for $q>0$ ($q<0$) lie below (above) with respect to the Schwarzschild case.

The energy per unit mass $E^*$ of the test particles is shown in Fig.~\ref{fig:energy} as a function of the normalized radial coordinate $r/M$ in the QM metric. The curves for fixed $q=0$ and arbitrary $j$ are presented in the left panel, while the curves for fixed $j=0$ and arbitrary $q$ are shown in the right panel. The left panel clearly shows that the energy of the particles for $j>0$ ($j<0$) cases is lower (higher) than in the Schwarzschild case.  In the right panel, the curve of $E^*$ for the Schwarzschild case is below (above) the curves of $q<0$ ($q>0$) cases.

The flux distribution as a function of the normalized radial coordinate $r/M$  is shown in Figs. (~\ref{fig:flux},~\ref{fig:fluxisco},~\ref{fig:fluxM}), which offers information on the amount of radiation emitted by the disk.

The left panel of Fig.~\ref{fig:flux} represents the scenario where $q=0$ is fixed and $j$ is arbitrary. In the full range of the radial distance, the flux for the Schwarzschild BH is consistently lower (higher) than the flux for the QM metric with $j<0$ ($j>0$), respectively. In the right panel, the flux for the Schwarzschild BH is always greater (lesser)  than the flux with $q<0$ ($q>0$). From these figures, one can clearly observe that the ER metric can mimic the physical characteristics of the Kerr BH.

We investigate whether an accretion disk in the ER spacetime can emulate the one predicted in the Kerr metric. As shown in the left panel of Fig.~\ref{fig:fluxisco}, the flux of the accretion disk in the ER metric for both positive and negative $q$ closely resembles the flux in the Kerr metric for certain positive and negative values of $j$. The flux in the ER metric was computed by fixing $q$, after which we numerically identified the maximum flux in the Kerr metric that matched the maximum flux in the ER metric. From this condition, we determined the corresponding values of $j$ in the Kerr metric. Afterwards, we calculated the values of $q$ and $j$ for the ER and Kerr metrics so that their flux maxima were identical. Even in these cases, while the maximum fluxes are the same, their radial positions relative to $r/M$ are shifted. Nevertheless, the mimicking effect remains evident.

The right panel of Fig.~\ref{fig:fluxisco} shows the flux in the Kerr and ER spacetimes for fixed $r_{isco}$. Here we first calculated the $r_{ISCO}$ in the ER spacetime for fixed $q=-4,-2,0,2,4$, then from the condition $r_{ISCO}(ER)=r_{ISCO}(Kerr)$ we found the values of $j$ and calculated the flux in the Kerr spacetime. It can be seen that the flux for the two spacetimes, as expected, with negative values of $j$ in the Kerr metric and $q$ in the ER metric are located below the Schwarzschild case, and positive values of $j$ and $q$ are located above, respectively.

Fig.~\ref{fig:fluxM}(left panel) shows the flux corresponding to different values of $q$ determined at fixed values of $j$ using the condition $M_2=0$ in the QM spacetime (see in Tab. \ref{tab:qjM}). This condition implies that rotating NSs, unlike Kerr BHs, can possess zero quadrupole moment. Hence, rotating NSs can reveal themselves only due to $J_1$, $J_3$ and higher multipole moments.

In the right panel of Fig.~\ref{fig:fluxM}, the flux is calculated at fixed values of $j$ and $M_{2}$ consequently for different values of $q_{1}$, $q_{2}$, and $q_{3}$ of the QM metric. If in the Kerr spacetime fixed $M_{2}$ implies fixed $j$, in the rotating NSs spacetime fixed $M_{2}$ does not imply that $j$ is fixed, it may also vary. In this sense, in addition to the mimicking effect, we have one more degree of freedom, i.e. $q$.

\begin{figure*}[ht]
\begin{minipage}{0.49\linewidth}
\center{\includegraphics[width=0.97\linewidth]{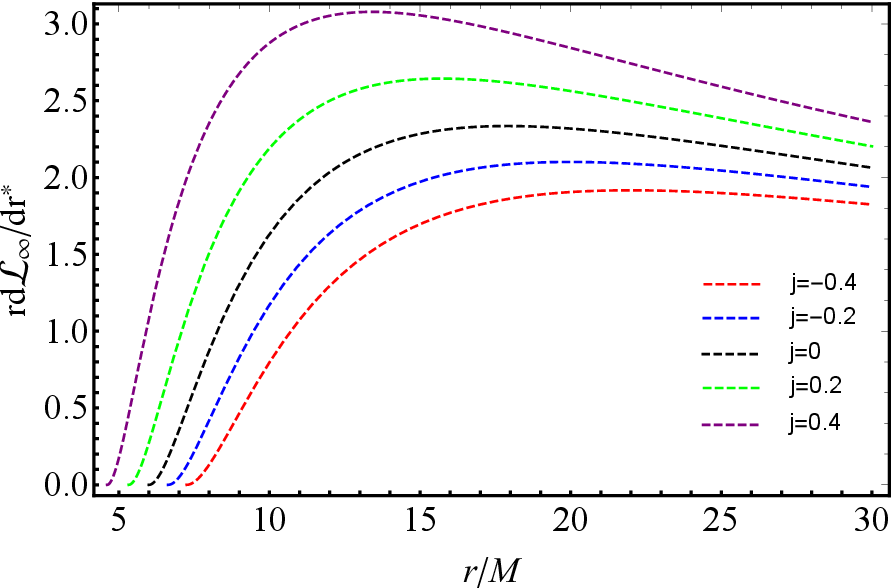}\\ } 
\end{minipage}
\hfill 
\begin{minipage}{0.50\linewidth}
\center{\includegraphics[width=0.97\linewidth]{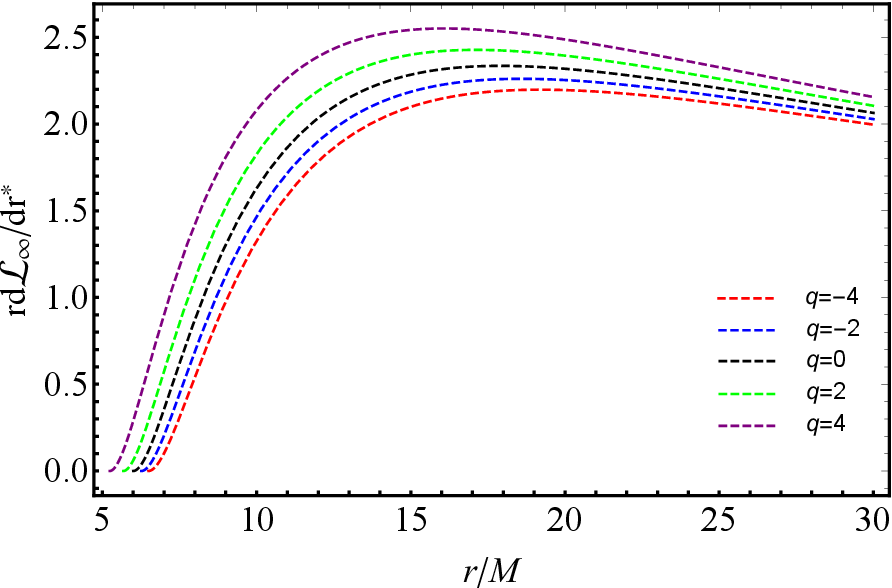}\\ }
\end{minipage}
\caption{Left panel: Differential luminosity in the oblate QM  metric with $q=0$. Right panel: Differential luminosity in QM metric with $j=0$.}
\label{fig:difflum}
\end{figure*}
In Fig.~\ref{fig:difflum}, we show the plot of the differential luminosity (the energy per unit of time that an observer receives at infinity) in the QM metric as a function of the normalized radial coordinate. The case of fixed $q=0$ and arbitrary $j$ is depicted in the left panel, whereas the case of fixed $j=0$ and arbitrary $q\geq 0$ is shown in the right panel. Due to the fact that differential luminosity is explicitly defined in terms of flux, we see that the behavior seen in Fig.~\ref{fig:flux}  readily transfers into the behavior of differential luminosity.

The differential luminosity in the Kerr spacetime for positive (negative) values of $j$ is larger (smaller) than in the Schwarzschild spacetime over the radial distance range, as shown in the left panel of ~\ref{fig:difflum}, which is consistent with the figures shown in the left panel of Fig. ~\ref{fig:flux}. Similarly, the trend seen in the right panel of Fig. ~\ref{fig:flux} is reflected in the right panel of Fig.~\ref{fig:difflum}.  Here, we compare the differential luminosity for the ER metric ($j=0$ and $q\geq 0$), to that of the Schwarzschild metric (black dashed curve). These results demonstrate the relationship between flux and differential luminosity, demonstrating that differential luminosity appropriately reflects the pattern seen in Fig. ~\ref{fig:flux}.

\begin{figure*}[ht]
\begin{minipage}{0.49\linewidth}
\center{\includegraphics[width=0.97\linewidth]{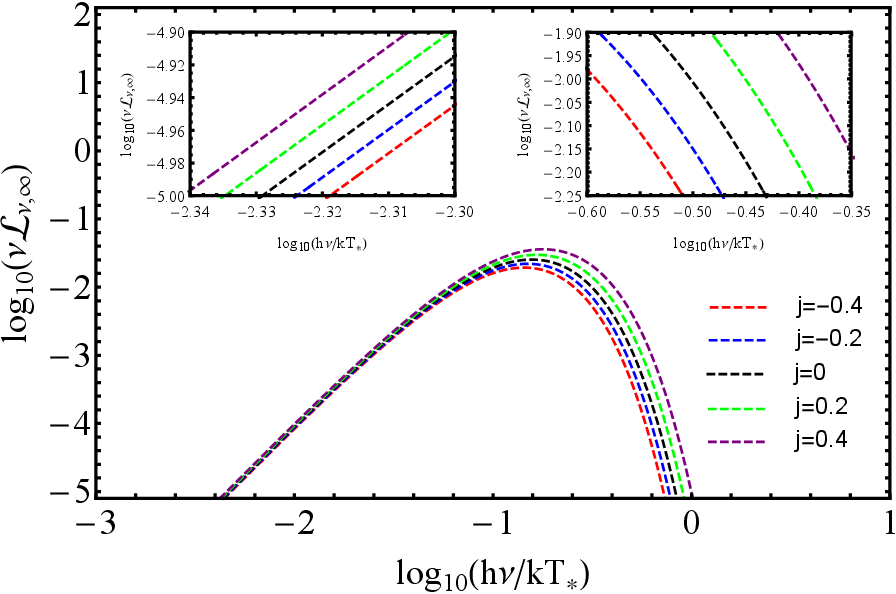}\\ } 
\end{minipage}
\hfill 
\begin{minipage}{0.50\linewidth}
\center{\includegraphics[width=0.97\linewidth]{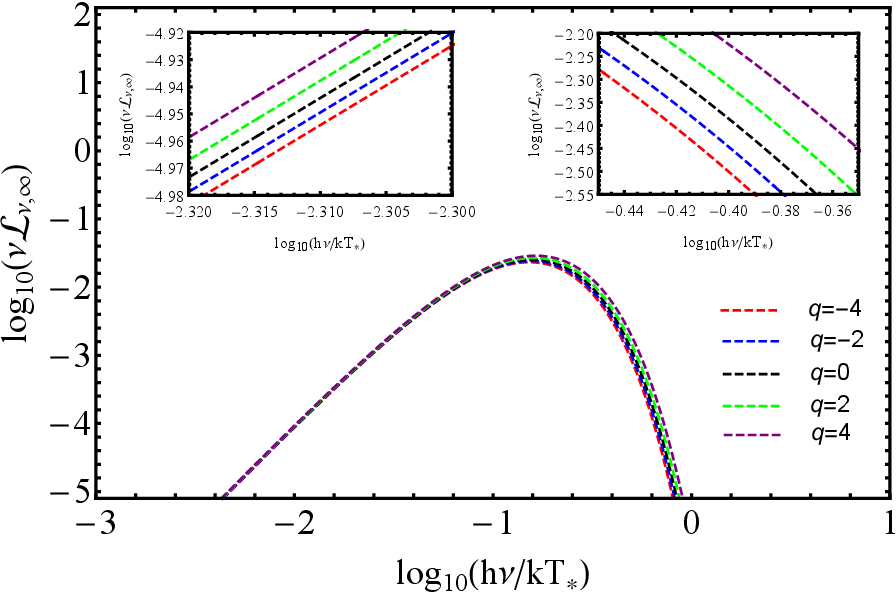}\\ }
\end{minipage}
\caption{Left panel: Spectral luminosity in the oblate QM metric with $q=0$. Right panel: Spectral luminosity in the oblate QM metric with $j=0$.}
\label{fig:lum_qj}
\end{figure*}

Finally, in Fig. ~\ref{fig:lum_qj}, we depict the spectral luminosity $\mathcal{L}_{\nu,\infty}$, in the QM metric, as a function of the frequency of radiation released by the accretion disk. The case of fixed $q=0$ and $j\geq0$ (Kerr metric) is represented in the left panel, while the case of fixed $j=0$ and $q\geq0$ (ER metric) is shown in the right panel.

\section{Final outlooks and perspectives}\label{sez5}

In this study we examined the features of thin accretion disks and the behavior of test particles in the gravitational field of spinning naked singularities represented by the QM spacetime. The distinguishing characteristics of naked singularities compared to Kerr BHs were highlighted by the analysis of orbital parameters, including angular velocity, energy, and angular momentum, as well as radiative flux, differential luminosity, and spectral luminosity.

Estimated disk features were shown to be strongly impacted by variations in quadrupole and spin parameters, indicating that the QM metric could be the effective means of evaluating various spacetime geometries through astrophysical observations.

%%%%%%%%%%%%%%%
We showed that the flux of the accretion disk within the ER metric, for both positive and negative values of $q$, mimics the flux observed in the Kerr metric for certain positive and negative values of $j$.

Furthermore, the flux for Kerr and ER metrics was analyzed under the condition of fixed $r_{ISCO}$. Initially, $r_{ISCO}$ was determined in the ER spacetime for specific values of $q$.  By equating $r_{ISCO}$ in both spacetimes, the corresponding values of $j$ were found in the Kerr metric, and the flux was subsequently calculated. The results showed that for negative values of $j$ in the Kerr metric and negative values of $q$ in the ER metric, the flux remained lower than in the Schwarzschild case, whereas positive values of these parameters led to higher flux.
%%%%%%%%%%%%%%%%%

Additionally, a flux distribution was examined under fixed multipole moment conditions, revealing how different values of the parameter influenced the accretion disk properties. The results demonstrated that, under specific conditions, the ER spacetime could serve as an effective model for studying the dynamics of the accretion disk.

\begin{acknowledgments}
The authors extend their sincere gratitude to  Hernando Quevedo for discussions on several conceptual aspects of this work. YeK acknowledges Grant No. AP19575366, TK acknowledges Grant No. AP19174979, KB and AU acknowledge Grant No. BR21881941 from the Science Committee of the Ministry of Science and Higher Education of the Republic of Kazakhstan. OL acknowledges the support by the  Fondazione  ICSC, Spoke 3 Astrophysics and Cosmos Observations. National Recovery and Resilience Plan (Piano Nazionale di Ripresa e Resilienza, PNRR) Project ID $CN00000013$ ``Italian Research Center on  High-Performance Computing, Big Data and Quantum Computing" funded by MUR Missione 4 Componente 2 Investimento 1.4: Potenziamento strutture di ricerca e creazione di ``campioni nazionali di R\&S (M4C2-19 )" - Next Generation EU (NGEU).

\end{acknowledgments}

%%%%%%%%%%%%%%%%%%%%%%%%%%%%%%%%%%%%%%%%%%%%%%%%%%%%%%%%%%%%%%%%%%%%%%%%%%%%%%%%%%%%%%%%%%%%%%%%%%%%%%%%%%%%%%%%%%%%%%%%%%%%%%%%%%%%%%%%%%%%%%%%%%%%%%%%%%%%%%%%%%%%%%%%%%%%%%%%%%%%%

\section{Appendix}\label{app}

In the main body of the text, we primarily compared the properties of accretion disks in the Kerr and ER spacetimes separately, even though both are limiting cases of the QM solution. Here, for completeness, we analyze how different values of $j$ and $q$ jointly influence accretion disk properties. To this end, we illustrate the angular velocity, angular momentum, and energy per unit mass for test particles in circular orbits, as well as the radiative flux, differential luminosity, and spectral luminosity in the QM spacetime.

\begin{figure*}[ht]
\begin{minipage}{0.49\linewidth}
\center{\includegraphics[width=0.97\linewidth]{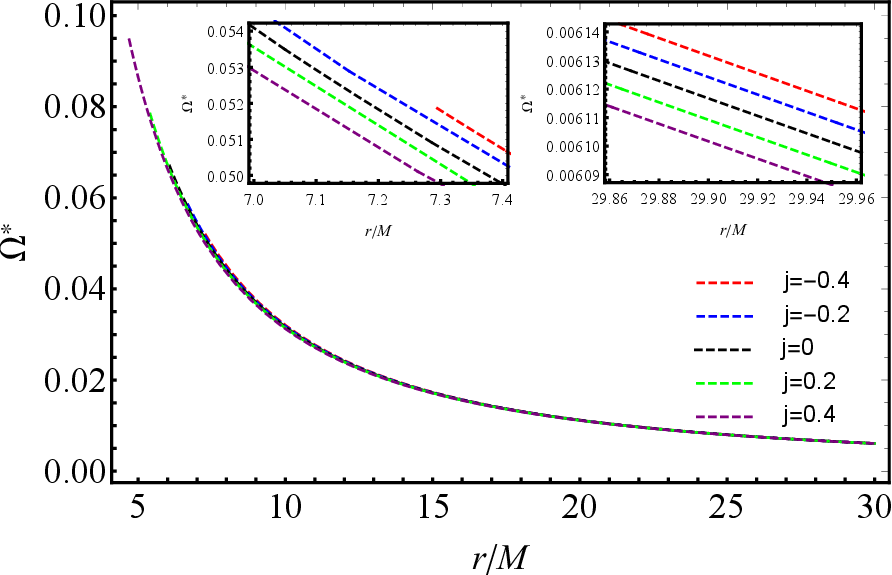}\\ }
\end{minipage}
\hfill 
\begin{minipage}{0.50\linewidth}
\center{\includegraphics[width=0.97\linewidth]{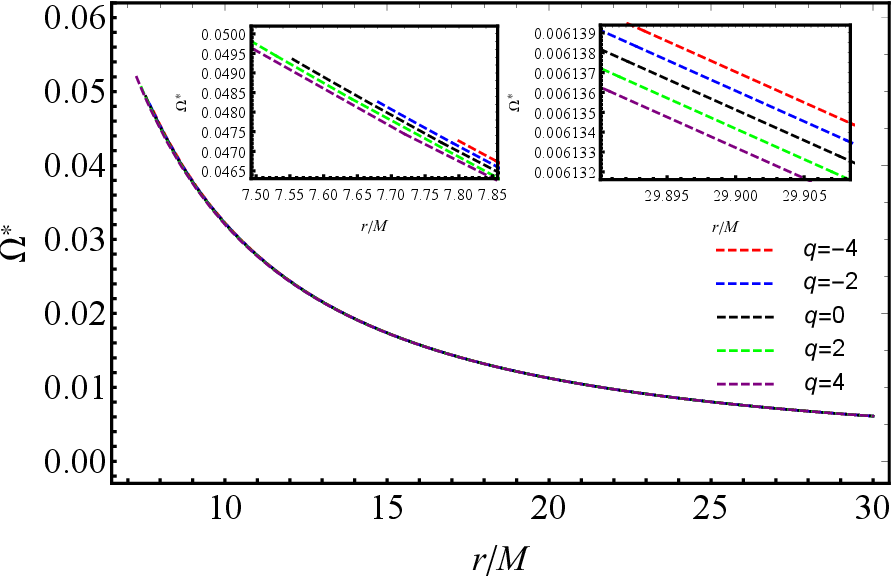}\\ }
\end{minipage}
\caption{Left panel: Angular velocity of test particles versus normalized radial distance $r/M$ in the QM metric with $q=-0.5$. Right panel: Angular velocity of test particles versus normalized radial distance $r/M$ in the QM metric with $j=-0.5$.}
\label{fig:Aomega}
\end{figure*}

In Fig. \ref{fig:Aomega} we show the orbital angular velocity $\Omega^*(r)$ of test particles as a function of the normalized radial coordinate $r/M$ in QM spacetime with different values of $j=\left[-0.4,-0.2,0,0.2,0.4\right]$ (left panel) and with different values of $q=\left[-4, -2, 0, 2, 4\right]$ (right panel). We clearly observe that when $j$ and $q$  have the same sign, their effects reinforce each other, whereas when they have opposite signs, depending on numerical values, their effects diminish, alleviate, or compensate for one other.

\begin{figure*}[ht]
\begin{minipage}{0.49\linewidth}
\center{\includegraphics[width=0.97\linewidth]{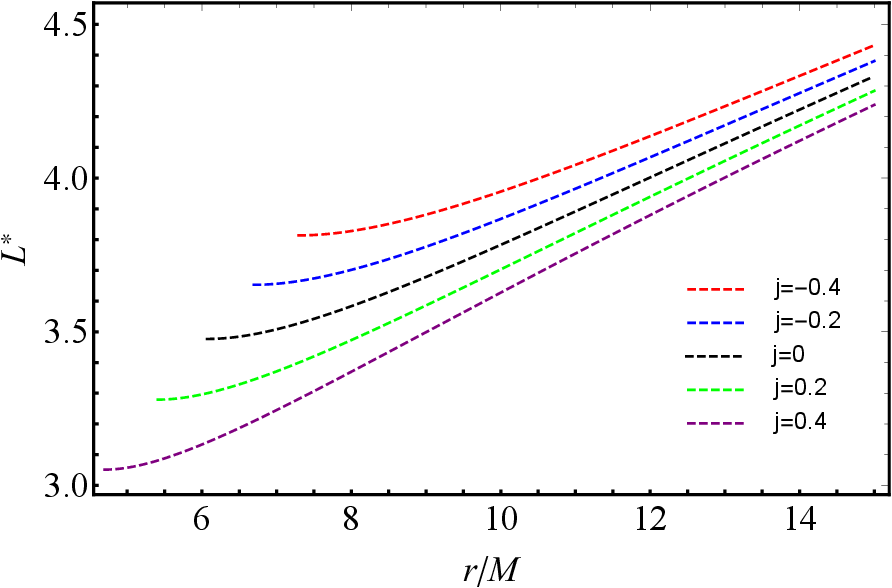}\\ } 
\end{minipage}
\hfill 
\begin{minipage}{0.50\linewidth}
\center{\includegraphics[width=0.97\linewidth]{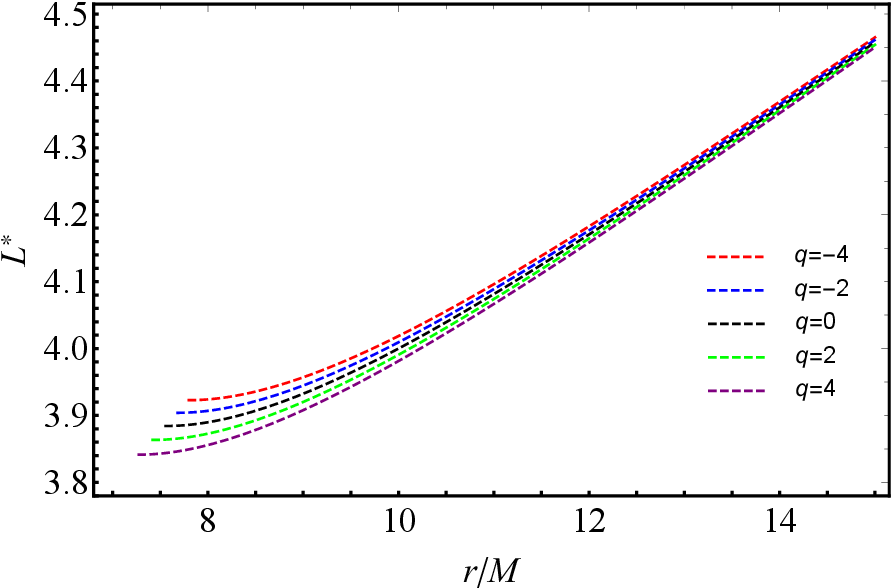}\\ }
\end{minipage}
\caption{Angular momentum $L^*$ of test particles versus normalized radial distance $r/M$ in the QM spactime. Left panel: $L^*$ for $q=-0.5$ and different values of $j$. Right panel: $L^*$ for $j=-0.5$ and different values of $q$.}
\label{fig:Aangmom}
\end{figure*}

In Fig. \ref{fig:Aangmom} we constructed the dimensionless orbital angular momentum $L^*(r)$ of test particles as
a function of the normalized radial coordinate $r/M$ for the QM metric, i.e. $q=-0.5$, with different values $j$ (left panel) and we fixed $j=-0.5$ and vary $q$ (right panel). Here we see that $L^*(r)$ is more sensitive to $j$ than to $q$.

\begin{figure*}[ht]
\begin{minipage}{0.49\linewidth}
\center{\includegraphics[width=0.97\linewidth]{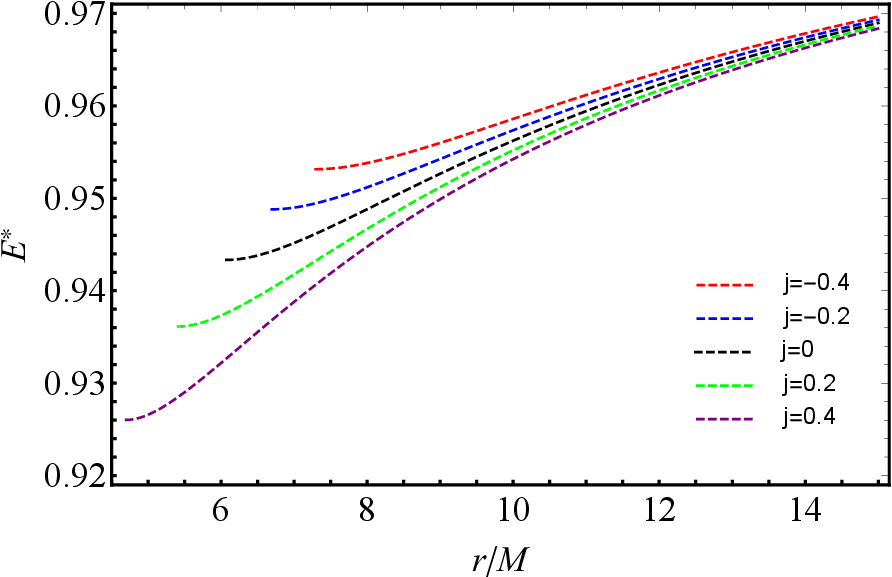}\\ } 
\end{minipage}
\hfill 
\begin{minipage}{0.50\linewidth}
\center{\includegraphics[width=0.97\linewidth]{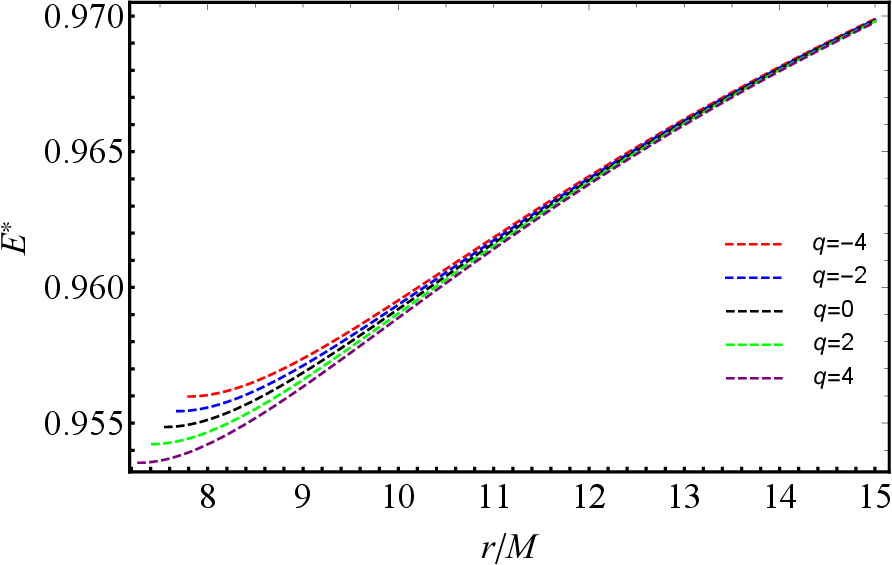}\\ }
\end{minipage}
\caption{Energy $E^*$ of test particles versus normalized radial distance $r$. Left panel: $E^*$ for $q=-0.5$ and different values of $j$. Right panel: $E^*$ for $j=-0.5$ and different values of $q$.}
\label{fig:Aenergy}
\end{figure*}

In Fig. ~\ref{fig:Aenergy}, we construct the energy per unit mass $E^*(r)$ of test particles as a function of the normalized radial coordinate $r/M$ in the QM spacetime. The left panel corresponds to fixed $q=-0.5$ and arbitrary $j$, while
the right panel represents fixed $j=-0.5$ and variable $q$. Here we see again that $j$ and $q$ with the same signs enhance each others effects, whereas  $j$ and $q$ with different signs diminish each other.

\begin{figure*}[ht]
\begin{minipage}{0.49\linewidth}
\center{\includegraphics[width=0.97\linewidth]{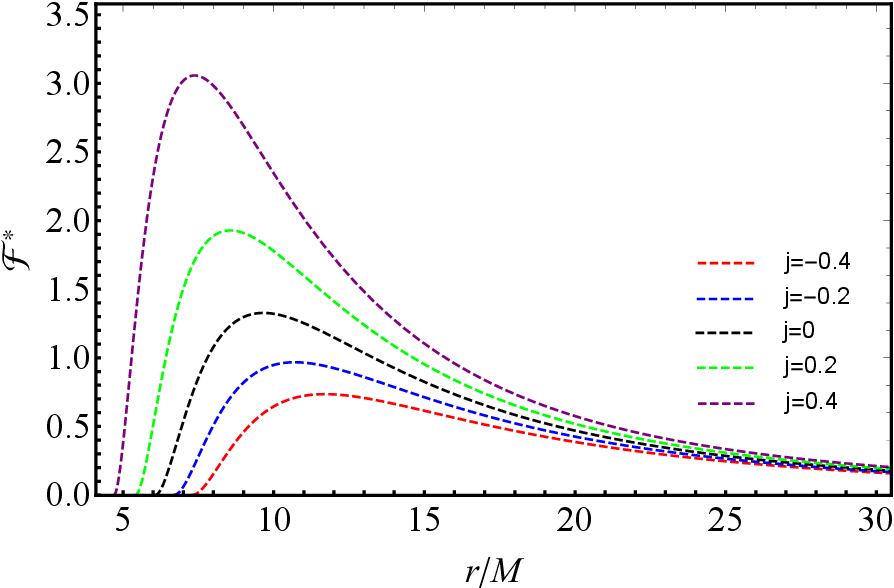}\\ } 
\end{minipage}
\hfill 
\begin{minipage}{0.50\linewidth}
\center{\includegraphics[width=0.97\linewidth]{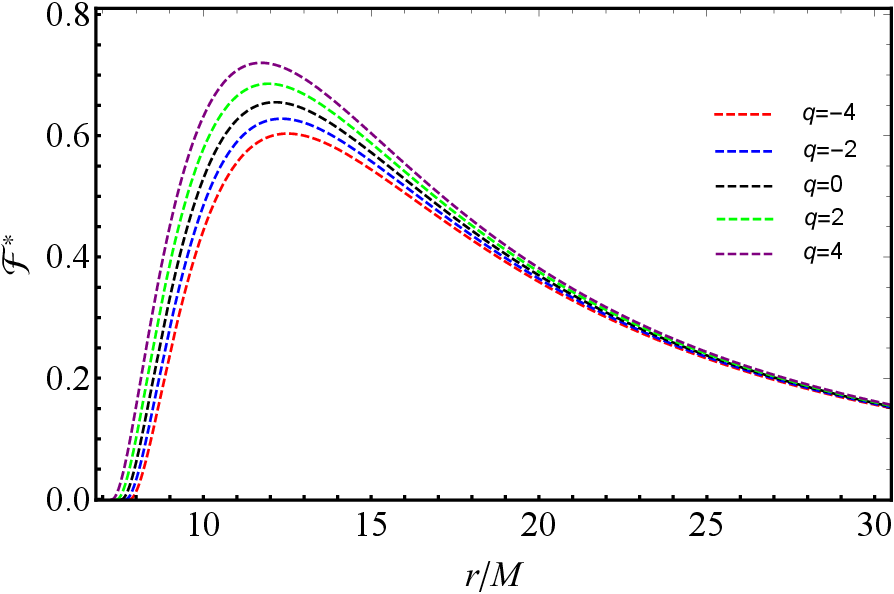}\\ }
\end{minipage}
\caption{Radiative flux $\mathcal{F}^*$ multiplied by $10^5$ of the accretion disk versus radial distance $r$ normalized in units of total mass $M$. Left panel: $\mathcal{F}^*$ for $q=-0.5$ and different $j$. Right panel: $\mathcal{F}^*$ for $j=-0.5$ and different $q$.}
\label{fig:Aflux}
\end{figure*}

The flux distribution, shown in Fig. ~\ref{fig:Aflux}, reveals information about the quantity of radiation the disk emits as a function of the radial coordinate. The left panel represents fixed $q=-0.5$ and variable $j$, while the right panel corresponds to fixed $j =-0.5$ and arbitrary $q$. We note that the radiative flux first increases with radius and subsequently decreases. In distant areas, the curve variations are minor. In contrast, notable differences appear in the inner region of the thin accretion disk. 

\begin{figure*}[ht]
\begin{minipage}{0.49\linewidth}
\center{\includegraphics[width=0.97\linewidth]{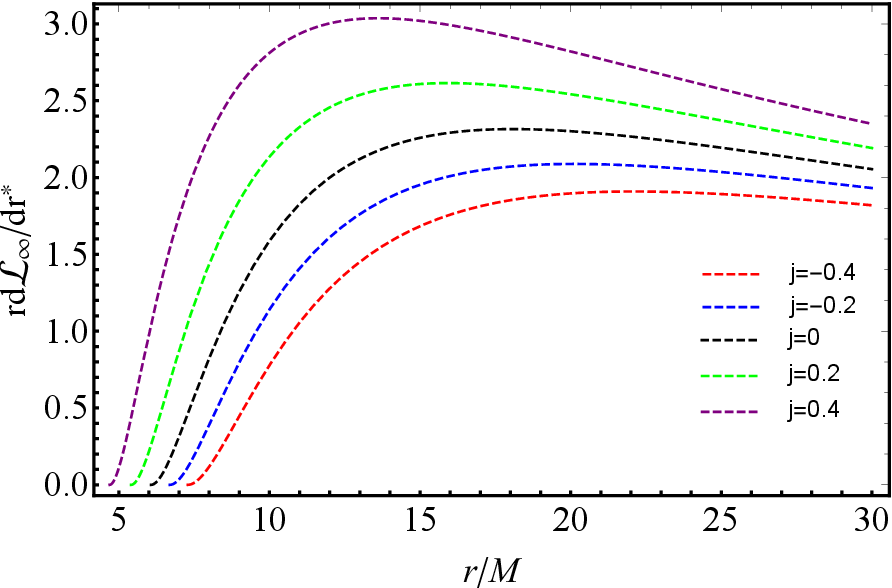}\\ } 
\end{minipage}
\hfill 
\begin{minipage}{0.50\linewidth}
\center{\includegraphics[width=0.97\linewidth]{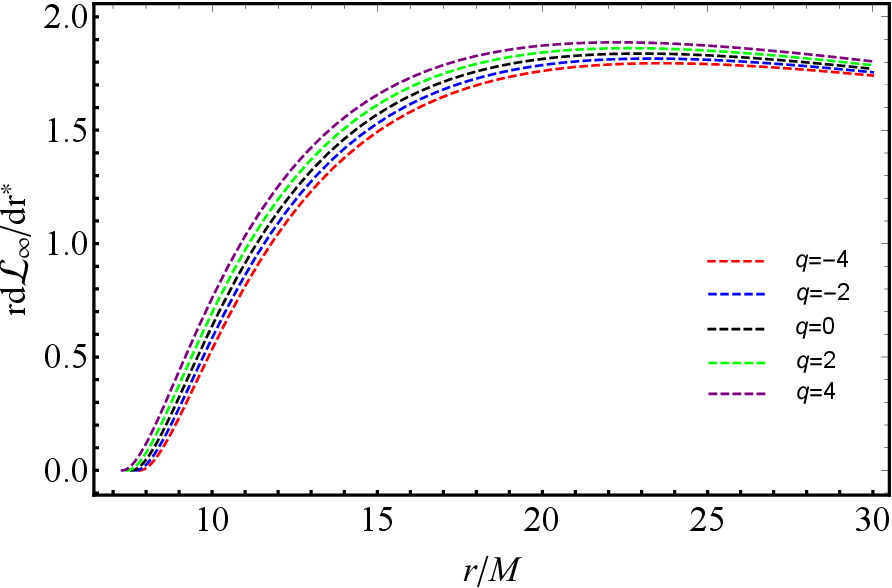}\\ }
\end{minipage}
\caption{Left panel: Differential luminosity for $q=-0.5$ and varying $j$. Right panel: Differential luminosity for $j=-0.5$ and varying $q$.}
\label{fig:Adifflum}
\end{figure*}

The differential luminosity as a function of the normalized radial coordinate for the QM is shown in Fig. ~\ref{fig:Adifflum}, where we set $q=-0.5$ and vary $j$ (left panel) and fix $j=-0.5$ and vary $q$ (right panel). As one can notice, the differential luminosity behaves similarly to the radiative flux illustrated in Fig.~\ref{fig:Aflux}. It is connected to the fact that Eq. ~\ref{eq:difflum} connects both quantities. As a result, everything seen in the flux is instantly translated into differential luminosity.

\begin{figure*}[ht]
\begin{minipage}{0.49\linewidth}
\center{\includegraphics[width=0.97\linewidth]{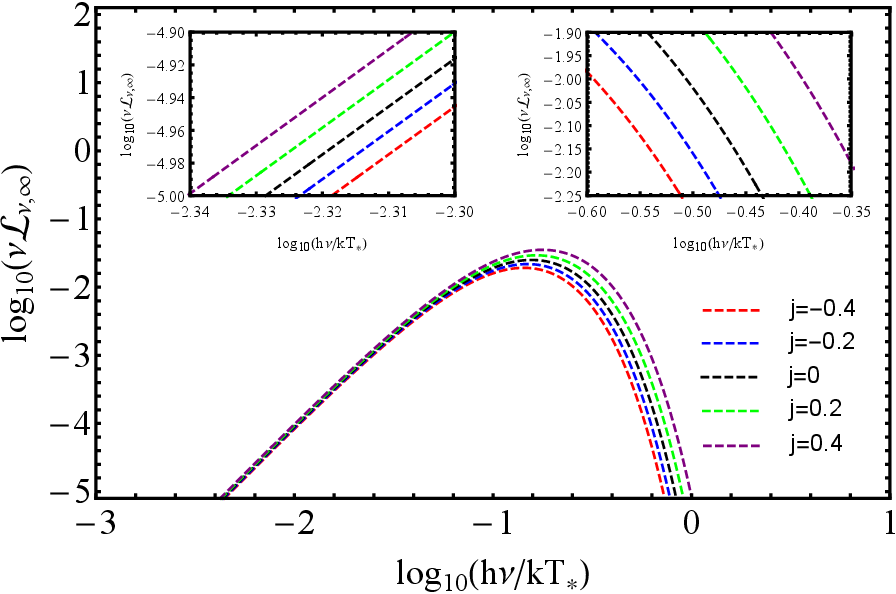}\\ } 
\end{minipage}
\hfill 
\begin{minipage}{0.50\linewidth}
\center{\includegraphics[width=0.97\linewidth]{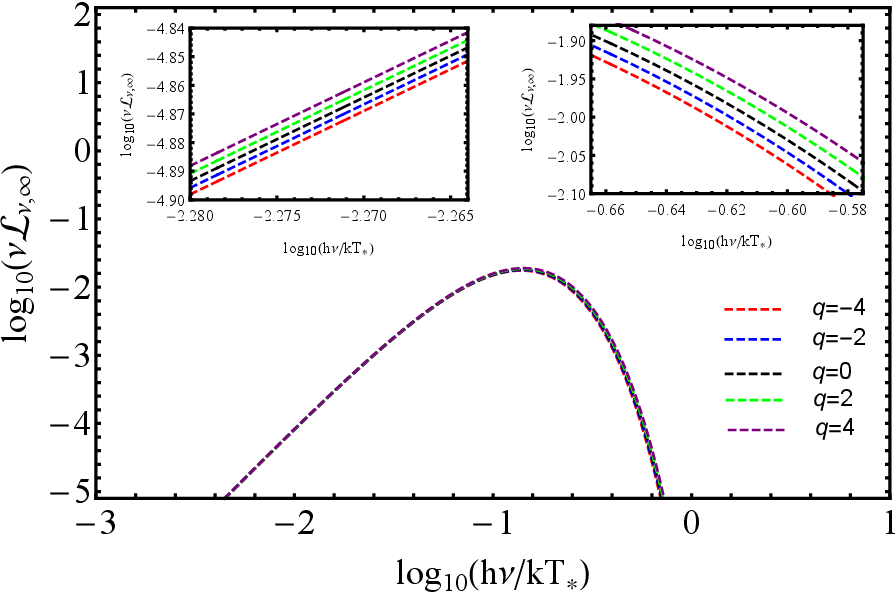}\\ }
\end{minipage}
\caption{Left panel: Spectral luminosity for $q=-0.5$ and different $j$. Right panel: Spectral luminosity for $j=-0.5$ and different $q$.}
\label{fig:Alum_qj}
\end{figure*}

The spectral luminosity distribution is shown as a function of radiation frequency in Fig. ~\ref{fig:Alum_qj}. As observed, all key characteristics of the accretion disk are more sensitive to $j$ than to $q$, and the mimicking effect is evident across all quantities.

\bibliography{0biblio}

\end{document}